\documentstyle[prl,aps,twoside,multicol]{revtex}

\begin{document}
\title{Reducing Constraints on Quantum Computer Design by Encoded Selective
Recoupling}
\author{D.A. Lidar and L.-A. Wu}
\address{Chemical Physics Theory Group, University of Toronto, 80 St. George
Str., Toronto, Ontario M5S 3H6, Canada}
\maketitle

\begin{abstract}
The requirement of performing both single-qubit and two-qubit operations in the
implementation of universal quantum logic often leads to very demanding
constraints on quantum computer design. We show here how to eliminate the
need for single-qubit operations in a large subset of quantum computer
proposals: those governed by isotropic and $XXZ,XY$-type anisotropic exchange
interactions. Our method employs an encoding of one logical qubit into
two physical qubits, while logic operations are performed using an
analogue of the NMR selective recoupling method.
\end{abstract}


\begin{multicols}{2}

Most proposals for quantum computer (QC) design rely on the execution
of both single-qubit and two-qubit operations. Typically these two types of
operations involve rather different manipulations and constraints.
This often leads to serious technical difficulties, a problem which
has been recognized and addressed in the context of QC proposals with
isotropic Heisenberg spin-exchange interactions
\cite{Loss:98,Kane:98Vrijen:00}, through the use of quantum codes
\cite{Bacon:99a,Kempe:00,Viola:00a,DiVincenzo:00a,Bacon:Sydney,Levy:01,Benjamin:01}.
It is important to note that isotropic exchange is an idealization
which in reality is likely to be perturbed due to surface and
interface effects, as well as spin-orbit coupling
\cite{Kavokin:01}. We focus here primarily on QC proposals that are
governed by anisotropic exchange interactions
\cite{Mozyrsky:01,Imamoglu:99,Quiroga:99,Zheng:00,Platzman:99}
(the $XXZ$ and $XY$ models, defined below).  These systems either
share some of the difficulties in implementing single qubit gates
exhibited by Heisenberg systems \cite{Loss:98,Kane:98Vrijen:00} (e.g.,
in the case of the quantum Hall proposal \cite{Mozyrsky:01}, extreme
$g$-factor engineering, highly localized and inhomogeneous magnetic
fields, and heating due to the high-intensity RF field needed for
single-spin operations), or have other problems resulting from the
need to implement both single and two qubit gates. E.g., in the
quantum dots in cavities proposal \cite{Imamoglu:99} elimination of
the single-qubit operations would halve the number of lasers,
significantly simplifying the experimental setup. There is therefore a
compelling motivation to re-examine the need for single-qubit
operations in the execution of quantum logic. Here we show how the
complications associated with single-qubit operations can be avoided
through the use of NMR-like recoupling methods \cite{Waugh:68Slichter:book}
applied to an encoding of two physical qubits into one logical qubit,
which we developed in \cite{WuLidar:01}. Our method allows universal
quantum logic to be attained through switching on/off exchange
interactions only, without requiring external single-qubit
operations. The method requires a modest overhead in the number of
physical qubits and gate operations, but this seems like a fair price
to pay in return for the reduction in complexity of experimental
setup. Through the use of recoupling we are able to unify the
treatment of both isotropic and anisotropic exchange. Similar to
\cite{Levy:01,Benjamin:01} but using less stringent methods, we show
how to reduce the overhead in the encoding proposed earlier for the
isotropic case \cite{Bacon:99a,Kempe:00,Viola:00a,DiVincenzo:00a} from
three physical qubits per logical (encoded) qubit to two, under the
assumption that the single-particle spectrum is non-degenerate. In the
$XXZ$ case our implementation of the controlled-phase ({\sc CPHASE})
gate uses as few as 4 interactions (in parallel mode on qubits
arranged in 1D). The efficient encoding, and the small overhead in
number of gate applications we report here, suggests that the hurdle
of single-qubit operations may be overcome in forthcoming experiments
implementing elementary quantum logic in isotropic or anisotropic
condensed or gas phase systems
\cite{Loss:98,Kane:98Vrijen:00,Mozyrsky:01,Imamoglu:99,Quiroga:99,Zheng:00,Platzman:99}.

{\it Exchange Hamiltonians.---} Using spin notation, the exchange
interaction quite generally \cite{comment1} has the form $H_{{\rm ex}}=\sum_{\alpha
=x,y,z}\sum_{i<j}J_{ij}^{\alpha }\sigma _{i}^{\alpha }\sigma
_{j}^{\alpha }$, where $\sigma _{i}^{\alpha }$ are the Pauli matrices,
and the summation is
over all qubit pairs $i,j$. Tunability of the exchange constants
$J_{ij}^{\alpha }$ is at the heart of all solid-state proposals, and
has been studied in detail, e.g., in \cite{Loss:98}. The isotropic (Heisenberg) case corresponds to $
J_{ij}^{\alpha}\equiv J_{ij}$. The $XY$ model is the case $J_{ij}^{x}=J_{ij}^{y}$,
$J_{ij}^{z}=0$. Examples
of QC proposals that fall into this category are: the quantum
Hall proposal \cite{Mozyrsky:01}, quantum dots \cite{Imamoglu:99,Quiroga:99} and
atoms in cavities \cite{Zheng:00}.
The $XXZ$ model is the case $J_{ij}^{x}=\pm J_{ij}^{y}\neq
J_{ij}^{z}$. [We refer to $+$ ($-$) as the axially symmetric
(antisymmetric) case.] When surface and interface effects are taken
into account the $XY$-examples, as well as the Heisenberg examples \cite{Loss:98,Kane:98Vrijen:00}, are better described by the
axially symmetric $XXZ$ model. Additional sources of non-zero
$J_{ij}^{z}$ in the $XY$-examples can be second-order effects (e.g., virtual cavity-photon
generation without spin-flips in \cite{Imamoglu:99}). A natural $XXZ$-example is that of
electrons on helium \cite{Platzman:99}. All these QC proposals were originally supplemented with 
{\em external} single qubit operations, which can be written as $
F=\sum_{i}f_{i}^{x}\sigma _{i}^{x}+f_{i}^{y}\sigma _{i}^{y}$. As argued
above, these operations almost invariably lead to various (system-specific)
difficulties, so we will not assume that they are available. In general one
must also consider the free Hamiltonian $H_{0}=\sum_{i}\frac{1}{2}
\varepsilon _{i}\sigma _{i}^{z}$, where $\varepsilon _{i}$ is the
single-particle spectrum. In general this spectrum will be
non-degenerate, e.g., due to different local $g$-factors \cite{Loss:98,Kane:98Vrijen:00}. Which of the {\em internal} parameters $
\{J_{ij}^{\alpha },\varepsilon _{i}\}$ are controllable is a system-specific
question, as summarized in Table 1. We now proceed to show how to perform
(encoded) universal quantum computation while respecting the constraints
imposed upon controllability of $\{J_{ij}^{\alpha },\varepsilon _{i}\}$ by
the various systems.

{\it Encoding and operations.---} First, we rewrite the general exchange
Hamiltonian in a form which emphasizes axial symmetry: 
\begin{equation}
H_{{\rm ex}}=\sum_{i<j}J_{ij}^{-}R_{ij}^{x}+J_{ij}^{+}T_{ij}^{x}+J_{ij}^{z}
\sigma _{i}^{z}\sigma _{j}^{z},  \label{eq:Hex}
\end{equation}
where 
\begin{equation}
T_{ij}^{x}=\frac{1}{2}\left( \sigma _{i}^{x}\sigma _{j}^{x}+\sigma
_{i}^{y}\sigma _{j}^{y}\right) \text{, } R_{ij}^{x}=\frac{1}{2}\left(
\sigma _{i}^{x}\sigma _{j}^{x}-\sigma _{i}^{y}\sigma _{j}^{y}\right) ,
\label{eq:TR}
\end{equation}
and $J_{ij}^{\pm }=J_{ij}^{x}\pm J_{ij}^{y}.$ Thus the axially
symmetric (anti-symmetric)
case corresponds to $J_{ij}^{-}=0$ ($J_{ij}^{+}=0$). Note that $T_{ij}^{x}$ can also be
written as $\sigma _{i}^{+}\sigma _{j}^{-}+\sigma _{i}^{-}\sigma _{j}^{+}$
[where $\sigma _{i}^{\pm }=(\sigma _{i}^{x}\pm i\sigma _{i}^{y})/2$], i.e.,
resonant energy transfer between qubit pairs. An important example of this
is the F\"{o}rster process, whereby through a Coulomb interaction an exciton
hops between neighboring quantum dots that are sufficiently close. This has
been used to show that a variety of quantum information processing tasks,
such as the preparation of entangled states of excitons, can be performed in
coupled quantum dots \cite{Quiroga:99}. Our results apply to this scenario
as well. In the axially symmetric case our code is simply: $|0_{L}\rangle
_{m}=|\uparrow \rangle _{2m-1}\otimes |\downarrow \rangle _{2m}$ and $
|1_{L}\rangle _{m}=|\downarrow \rangle _{2m-1}\otimes |\uparrow \rangle
_{2m} $ for the $m^{{\rm th}}$ encoded qubit, $m=1,...,N/2$. In the axially
antisymmetric case: $|0_{L}\rangle =|\uparrow
\uparrow \rangle $ and $|1_{L}\rangle =|\downarrow \downarrow \rangle $
(in simplified notation). Thus, logical qubits
correspond to pairs of nearest neighbor physical qubits (e.g.,
spins). Preparation and measurement of these states was discussed in
Ref. \cite{WuLidar:01}. Briefly, preparation relies on relaxation to the
ground state of the Hamiltonian $T_{ij}^x$ ($R_{ij}^x$) in the axially
symmetric (antisymmetric) case, while measurement (which can also be
used for preparation) employs an analogue
of Kane's a.c. capacitance scheme \cite{Kane:98Vrijen:00}. It is important to note that
none of the terms in $H_{{\rm ex}}$ is capable of flipping spins $i,j$
{\em separately}. Therefore the axially symmetric and antisymmetric subspaces are
decoupled. This means that we can independently operate on the corresponding
subspaces. Our discussion below is carried out in tandem for these two
cases. With the single number $m$ serving to label our encoded qubits, it is
advantageous to compactify our notation further. Let $J_{m}^{\alpha }\equiv
J_{2m-1,2m}^{\alpha }$ ($\alpha =z,\pm $), and $\epsilon _{m}^{\pm }\equiv
\left( \varepsilon _{2m-1}\pm \varepsilon _{2m}\right) /2$.

Let us now introduce operators which implement rotations on the encoded
qubits. Let $T_{m}^{x}\equiv T_{2m-1,2m}^{x}$ and $T_{m}^{z}=\frac{1}{2}
\left( \sigma _{2m-1}^{z}-\sigma _{2m}^{z}\right) $; $R_{m}^{x}\equiv
R_{2m-1,2m}^{x}$ and $R_{m}^{z}=\frac{1}{2}\left( \sigma _{2m-1}^{z}+\sigma
_{2m}^{z}\right) $, where $T_{2m-1,2m}^{x}$ and $R_{2m-1,2m}^{x}$ were
defined in Eq.~(\ref{eq:TR}). As we showed in \cite{WuLidar:01}, $T_{m}^{x}$
($R_{m}^{x}$)\ acts as the Pauli $\sigma ^{x}$ matrix on the axially
symmetric (antisymmetric)\ $m^{{\rm th}}$ encoded qubit. Similarly, $T_{m}^{z}$ ($R_{m}^{z}$) acts as $\sigma ^{z}$. Therefore pairs of these
operators each generate an ``encoded $SU(2)$'' group on the logical qubits,
i.e., the group of all single-encoded-qubit operations. Moreover, $[T_m^\alpha,R_m^\beta]=0$
(in agreement with the decoupling
of the symmetric and antisymmetric subspaces), so that these
single-encoded-qubit operations can be implemented in classical parallelism. Let us now
momentarily assume that we have independent control over all parameters ($
\epsilon _{m}^{\pm },J_{ij}^{\pm },J_{ij}^{z}$). Below we will relax this
constraint by using selective recoupling. In order to implement
single-encoded-qubit operations, we turn on the interactions $J_{m}^{\pm }$
and the energy sums and differences $\epsilon _{m}^{\pm }$, while
leaving off the interactions between spins belonging to different encoded qubits 
(i.e., $J_{2m,2m+1}^{\alpha }=0$), as well as leaving all $J_{ij}^{z}$ off.
We can then rewrite the total Hamiltonian $H=H_{0}+H_{{\rm ex}}$ as: 
\begin{equation}
H=\sum_{m=1}^{N/2}\left( \epsilon
_{m}^{-}T_{m}^{z}+J_{m}^{+}T_{m}^{x}\right) +\left( \epsilon
_{m}^{+}R_{m}^{z}+J_{m}^{-}R_{m}^{x}\right) ,  \label{eq:H1}
\end{equation}
while omitting a constant term. Written in this form, it is
clear that by selectively turning on/off the parameters $\epsilon
_{m}^{-},J_{m}^{+}$ ($\epsilon _{m}^{+},J_{m}^{-}$) for the axially
symmetric (antisymmetric) qubit, one can implement all single-encoded-qubit
operations, by using Euler angle rotations to generate the ``encoded $SU(2)$'' group. Moreover, $H$ is expressed as a sum over terms acting on different
encoded qubits, so that all $N/2$ encoded qubits (of a given symmetry) can
be operated on independently. In other words, the encoded Hilbert space has
a tensor product structure.

To complete the general discussion we must also show how to couple different
encoded qubits through a non-trivial (entangling) gate. For the $XXZ$ model
this turns out to be even simpler than implementing single-encoded-qubit
operations. For, turning on the coupling $J_{2m,2m+1}^{z}$ between pairs of
spins belonging to two neighboring encoded qubits immediately implements the
encoded $-T_{m}^{z}T_{m+1}^{z}$ ($R_{m}^{z}R_{m+1}^{z}$) Hamiltonian on the
axially symmetric (antisymmetric) qubits. To see this, note that in the
axially symmetric case, e.g., $|0_{L}\rangle_1 |0_{L}\rangle_2 =|01\rangle
_{12}|01\rangle _{34}\stackrel{\sigma _{2}^{z}\sigma _{3}^{z}}{\rightarrow }
-|01\rangle _{12}|01\rangle _{34}=-|0_{L}\rangle_1 |0_{L}\rangle_2 $, and
similarly for the other three combinations: $|0_{L}\rangle |1_{L}\rangle
\rightarrow |0_{L}\rangle |1_{L}\rangle $, $|1_{L}\rangle |0_{L}\rangle
\rightarrow |1_{L}\rangle |0_{L}\rangle $, $|1_{L}\rangle |1_{L}\rangle
\rightarrow -|1_{L}\rangle |1_{L}\rangle $ so that in all $\sigma
_{2}^{z}\sigma _{3}^{z}$ indeed acts as $-T_{1}^{z}T_{2}^{z}$. Since as is
well known \cite{Nielsen:book}, the controlled-phase ({\sc CPHASE}) gate is
directly obtainable by turning on the Hamiltonian $\sigma ^{z}\otimes \sigma
^{z}$ between physical qubits, in our case turning on $J_{2m,2m+1}^{z}$
yields a {\sc CPHASE} gate between encoded qubits. We consider the $XY$ model
below, since it requires the introduction of the selective recoupling
method. For the $XXZ$ model, this encoded-{\sc CPHASE} together with the
single-encoded-qubit operations suffices to perform encoded universal
quantum computation \cite{Barenco:95a}.

{\it Recoupling and Encoding Recoupling.--- }Recall that, as discussed
above, in each instance of $H_{{\rm ex}}$ (Heisenberg, $XY$, $XXZ$) one
typically has control over only one type of parameter out of the set $
\{J_{ij}^{\alpha },\varepsilon _{i}\}$. In our treatment above we made
liberal use of all parameters, and it is now time to relax this assumption.
To show this we now demonstrate how selective recoupling, applied to our
encoded qubits, provides the requisite flexibility. Let us first briefly
recall the basic idea of selective recoupling \cite{Waugh:68Slichter:book} through a
simple NMR example. In a 2-spin molecule in NMR, the internal Hamiltonian is 
$H_{{\rm NMR}}=\sum_{i=1}^{2}\varepsilon _{i}\sigma
_{i}^{z}+J_{12}^{z}\sigma _{1}^{z}\sigma _{2}^{z}$ with uncontrollable
parameters $\varepsilon _{i}$, $J_{12}^{z}$. However, control over $
\varepsilon _{i}$ is needed to implement $z$-axis rotations, while control
over $J_{12}^{z}$ is needed to implement a {\sc CPHASE}. This is done by
pulsing an external magnetic field along the $x$-axis. Let $A$ and $B$ be
anticommuting hermitian operators where $A^{2}=I$ ($I$ is the identity
matrix). Then the operation of ``conjugating by $A$'', 
\begin{eqnarray}
C_{A}\circ \exp (iB) &\equiv& \exp (-iA\pi /2)\exp (iB)\exp (iA\pi /2) 
\nonumber \\
&=& \exp(-iB)
\end{eqnarray}
causes $B$'s sign to be flipped. Thus: $\exp (-iH_{{\rm NMR}}\tau )\left[
C_{\sigma _{1}^{x}}\circ \exp (-iH_{{\rm NMR}}\tau )\right] =\exp [-2i\tau
\varepsilon _{2}\sigma _{2}^{z}]$, which implements a rotation through an
angle $\theta =2\tau \varepsilon _{2}$ about the $z$-axis of the second spin. Notably, the Ising coupling term $
\sigma _{1}^{z}\sigma _{2}^{z}$ has been eliminated. A similar calculation
reveals that: $\exp
(-iH_{{\rm NMR}}\tau )\left[ C_{\sigma _{2}^{x}}\circ C_{\sigma
_{1}^{x}}\circ \exp 
(-iH_{{\rm NMR}}\tau )\right] =\exp [2i\tau J_{12}^{z}\sigma _{1}^{z}\sigma
_{2}^{z}]$, i.e., the selective implementation of the Ising coupling term
through an angle $\tau J_{12}^{z}$ . Notice that to achieve this effect all
that was needed was control over the parameters turning on/off the $\sigma
_{i}^{x}$ terms. Physically, the reason that $H_{
{\rm NMR}}$ was neglected during the $\pi /2$ rotations is that typically in
NMR the $\sigma _{i}^{x}$ terms can be made much larger than $H_{{\rm NMR}}$. Selective recoupling methods can be extended to deal with any number of
spins coupled through an NMR-type Hamiltonian, and
efficient methods using Hadamard matrices have been developed for both
homonuclear \cite{Linden:99c} and heteronuclear systems \cite
{Leung:00,Jones:99}.

Consider now an $XXZ$-type Hamiltonian where the $J_{m}^{+}$ parameters are
controllable but $\epsilon _{m}^{-}$ and $J_{m}^{z}$ are fixed. As
argued above, this is a model of the $XY$-examples of QC proposals \cite{Mozyrsky:01,Imamoglu:99,Quiroga:99,Zheng:00}
that takes certain symmetry-breaking mechanisms into account. We can
now map the selective recoupling method directly onto our problem. For
simplicity let us consider just the axially symmetric case. Then we can
rewrite $H=H_{0}+H_{{\rm ex}}$ as: $H_{{\rm AX}}=\sum_{m=1}^{N/2}\epsilon
_{m}^{-}T_{m}^{z}-J_{m}^{z}T_{m}^{z}T_{m+1}^{z}+J_{m}^{+}T_{m}^{x}$, where
we have omitted a constant term. The
important point is now that $T_{m}^{x}$ and $T_{m}^{z}$ satisfy the
properties required of $A$ and $B$ above, {\em on the code subspace}. In fact, the structure of $H_{{\rm 
AX}}$ is exactly analogous to that of $H_{{\rm NMR}}$, the
only difference being that the $T$ operators act on encoded qubits as
opposed to directly on physical spins. Hence the argument that held for $H_{
{\rm NMR}}$ holds here as well: By using recoupling through ``conjugation
by $T_{m}^{x}$'' we can selectively turn on and off the single-encoded-qubit
rotation $T_{m}^{z}$ and the encoded-Ising interaction
$T_{m}^{z}T_{m+1}^{z}$. This example of ``encoded selective
recoupling'' establishes that encoded universal computation in the
$XXZ$ model can be done using control over the $J_{m}^{+}$ parameters alone.

Next, consider the $XY$ model, i.e., the idealized version of the proposals
in \cite{Mozyrsky:01,Imamoglu:99,Quiroga:99,Zheng:00}, with controllable $J_{ij}^{+}$, but fixed $\varepsilon _{i}$. We
still use the encoding $|0_{L}\rangle =|\uparrow \downarrow \rangle $, $
|1_{L}\rangle =|\downarrow \uparrow \rangle $. To
implement encoded single-qubit operations, we can use the same
encoded recoupling method as for the $XXZ$ model. As for encoded two-qubit operations, we now
no longer have the $\sigma _{i}^{z}\sigma _{j}^{z}$ terms. Since the
$XY$ model with nearest-neighbor interactions can be shown not to be
universal \cite{WuLidar:01}, we turn on also next-nearest neighbor $J_{ij}^{+}$ terms (these can
still be nearest-neighbor in a 2D hexagonal geometry). First
note that $C_{T_{12}^{x}}\circ T_{23}^{x}= i \sigma _{1}^{z} \sigma _{2}^{z}T_{13}^{x}$. Now
assume we can control $J_{13}^{+}$; then, using conjugation by $\pi /4$: $C_{
\frac{1}{2}T_{13}^{x}}\circ \left( C_{T_{12}^{x}}\circ T_{23}^{x}\right)
=\sigma _{2}^{z}(\sigma _{3}^{z}-\sigma _{1}^{z})/2$. Since $\sigma
_{1}^{z}\sigma _{2}^{z}$ is constant on the code subspace it can be ignored.
On the other hand, $\sigma _{2}^{z}\sigma _{3}^{z}$ again acts as $
-T_{1}^{z}T_{2}^{z}$, i.e., as an encoded $\sigma ^{z}\otimes \sigma ^{z}$.
This establishes universal encoded computation in the $XY$ model.

{\it Cost.---} Let us now count how many elementary steps are needed to
implement the various quantum computing primitives in the $XXZ$
model. We define such a step as a single pulse whereby a single
$J_{ij}^\alpha$ is switched on and then off. We
will only assume the least demanding architecture of nearest neighbor
interactions and a 1D layout of spins. Improvements are certainly possible
with next-nearest neighbor interactions and/or a 2D geometry. For
single-encoded-qubit operations, it takes one step to turn on a
rotation about the encoded $x$-axis (under the standard assumption that we
can make $|J_{m}^{+}|\gg |\epsilon _{m}^{-}|,|J_{m}^{z}|$), while it takes 4
steps to implement a rotation about the encoded $z$-axis (turn on $
T_{m}^{x}$ for $\pi /2$, free evolution under $\epsilon
_{m}^{-}T_{m}^{z}-J_{m}^{z}T_{m}^{z}T_{m+1}^{z}$, repeat with $T_{m}^{x}$
for $-\pi /2$). Therefore using the standard Euler angle construction it
takes at most $6$ steps to implement any single-encoded-qubit rotation.
The encoded {\sc CPHASE} operation is similar: If we assume that 
$T_{m}^{x}$ and $T_{m+1}^{x}$ can be turned on in parallel then the same
count of 4 steps as for encoded $z$-axis rotations applies; otherwise
we need to add 2 more
operations, for a total of 6 steps. In the $XY$ case the
implementation of {\sc CPHASE} given above takes 5 steps.

{\it Encoded recoupling for the Heisenberg case.---} Elimination of
single-qubit operations in the isotropic exchange case was first shown in 
\cite{Bacon:99a}, and further developed in \cite{Kempe:00,Viola:00a,DiVincenzo:00a}. With a fully degenerate $H_0$ this required encoding one qubit into three. We now show how to
simplify this encoding, assuming a non-degenerate $H_{0}$. Our code is the
same as for the axially symmetric case above, and the same as that used in 
\cite{Levy:01,Benjamin:01}, but our method is less stringent. Ref. \cite{Levy:01}
used spin-resonance techniques, with the rather demanding requirement that
the spin-spin interaction strength be modulated at high frequency. Ref. \cite
{Benjamin:01} showed how universal computation can be be performed by
varying the strength of the exchange coupling using non-oscillatory pulses.
This scheme is very much in the spirit of our solution, but it has the
problem of undesired spin-rotations taking place while the interaction is
off. We solve this problem here using the selective recoupling method. Let
us write the total Hamiltonian as $H=H_{0}+H_{{\rm Heis}}$, where $H_{{\rm 
Heis}}=\sum_{i<j}J_{ij}(T_{ij}^{x}+\frac{1}{2}\sigma _{i}^{z}\sigma _{j}^{z})
$ with $\frac{1}{2}J_{ij}^{+}=J_{ij}^{x}=J_{ij}^{y}=J_{ij}^{z}=J_{ij}.$ The
exchange coupling parameters $J_{ij}$ are assumed to be controllable, while $
H_{0}$ is not (except by application of a global magnetic field). Therefore, similar to the anisotropic case, when we turn on $
J_{m}\equiv J_{2m-1,2m}$ the Hamiltonian can be written as $H_{{\rm Heis}
}=\sum_{m=1}^{N/2}(\epsilon _{m}^{-}T_{m}^{z}+J_{m}T_{m}^{x})+\Delta $,
where $\Delta =\sum_{m=1}^{N/2}(\epsilon _{m}^{+}R_{m}^{z}+$ $\frac{J_{m}}{2}
\sigma _{2m-1}^{z}\sigma _{2m}^{z})$ acts trivially on the code space and
hence can be omitted. By recoupling using $T_{m}^{x}$ we can selectively
turn on and off the single-encoded-qubit rotation $T_{m}^{z}$, as above,
thus generating the encoded-$SU(2)$ group only on the desired encoded-qubit.
Next, let us show how to selectively turn on a two-qubit Hamiltonian such as 
$T_{1}^{z}T_{2}^{z}=\sigma _{2}^{z}\sigma _{3}^{z}$. If we directly turn on $
h_{23}=J_{23}\sum_{\alpha=x,y,z}\sigma _{2}^{\alpha}\sigma _{3}^{\alpha}$, the encoded space will leak.
Recoupling can extract the $\sigma _{2}^{z}\sigma _{3}^{z}$ term as follows.
First, note that $C_{T_{1}^{x}}\circ e^{-i\pi T_{1}^{z}/2}=e^{i\pi
T_{1}^{z}/2}$, and $C_{T_{1}^{z}}\circ h_{23} = J_{23}(-\sigma _{2}^{x}\sigma _{3}^{x}-\sigma _{2}^{y}\sigma
_{3}^{y}+\sigma _{2}^{z}\sigma _{3}^{z})$. Hence $e^{-ih_{23}t/2} C_{T_{1}^{z}}\circ e^{-ih_{23}t/2} = e^{-iJ_{23}\sigma_{2}^{z}\sigma _{3}^{z}t}$, so that $
T_{1}^{z}T_{2}^{z}$ may be implemented selectively using 6 steps,
which completes the 
requirements for universal computation. It is
interesting to contrast these results with the 19 steps required in
serial mode for the analogous operation in the
isotropic case, assuming fully degenerate $H_0$ (7 steps are required in parallel mode in 2D) \cite{DiVincenzo:00a}. As a final comment, note that the
code we used here is a decoherence-free subspace (DFS) and thus offers automatic protection against collective
dephasing errors \cite{Kempe:00}, as recently demonstrated in an ion
trap experiment \cite{Kielpinski:01}. When other errors are present
one may use the method of concatenating DFSs with quantum error
correcting codes \cite{Lidar:PRL99}, at the price of introducing
greater qubit overhead, or use a combination of recoupling and decoupling techniques \cite{ByrdLidar:01}.

{\it Conclusions.---} The requirement of performing both single- and
two-qubit operations in one quantum computing device often leads to severe
technical constraints and difficulties. We have shown here that selective
recoupling, as applied to encoded qubits, is a very general method to
overcome these problems. It allows all quantum logic operations to be
performed by turning on/off pairwise exchange interactions. The
trade-off is modest: a qubit is encoded into the state of two
neighboring spins, and universal quantum logic gates require only 4-6
interactions to be turned on/off in a simple 1D geometry with nearest
neighbor coupling. We believe that this alternative to the hard requirements
of quantum computing with single-qubit gates may substantially
simplify the design of quantum computers.

{\it Acknowledgements.---} DAL acknowledges support from PREA, PRO, the
Connaught Fund, and AFOSR (F49620-01-1-0468). We thank
Profs. V. Privman, A. Imamo$\bar{\rm g}$lu, and B.E. Kane for useful
correspondence.

\end{multicols}

\newpage

\label{tab1}
\begin{table}[tbp] \centering
\begin{tabular}{|c|c|c|c|}
\hline
System & Two-qubit Hamiltonian & $H_{0}=\sum_{i}\frac{1}{2}
\varepsilon _{i}\sigma _{i}^{z}$ & External $f_i^{x,y}$ 
\\ \hline
Spin-coupled quantum dots \cite{Loss:98} & Heisenberg, controllable &
fixed & hard
\\ 
Donor atom nuclear/electron spins \cite{Kane:98Vrijen:00} &
Heisenberg, controllable & fixed & hard
\\ 
Quantum Hall \cite{Mozyrsky:01} & $XY$, controllable & fixed & hard \\ 
Quantum dots/atoms in cavities
\cite{Imamoglu:99,Zheng:00} & $XY$,
controllable & controllable & easy, requires additional lasers \\ 
Exciton-coupled quantum dots \cite{Quiroga:99} & $XY$, controllable &
fixed & hard
\\ 
Electrons on helium \cite{Platzman:99} & $XXZ$, only $J_{ij}^+$
controllable & controllable & easy but slow and hard to tune
\\ \hline
\end{tabular}
\caption{Comparison of some QC proposals in terms of difficulty of
implementing two-qubit
($J_{ij}^\alpha$), internal ($\epsilon_i$), and external single-qubit
operations ($f_i^{x,y}$). $H_0=$ fixed means that it is hard to
independently change each $\varepsilon_i$. \label{key}}
\end{table}


\begin{thebibliography}{10}
\bibitem{comment1}
When spin-orbit interaction is taken into account this must be modified to
include cross terms \cite{Kavokin:01}. We will analyze this perturbation in
a forthcoming publication.
\end{thebibliography}
\end{document}